# Optical frequency comb integration in radio telescopes: advancing signal generation and phase calibration

Minji Hyun,[1,6] Changmin Ahn,[1,6] Junyong Choi,[1] Jihoon Baek,[1] Woosong Jeong,[1] Do-Heung Je,[2] Do-Young Byun,[2,3] Jan Wagner,[4] Myoung-Sun Heo,[5] Taehyun Jung[2] & Jungwon Kim[1,*]

[1] *Korea Advanced Institute of Science and Technology (KAIST), Daejeon 34141, Korea*
[2] *Korea Astronomy and Space Science Institute (KASI), Daejeon 34055, Korea*
[3] *University of Science and Technology (UST), Daejeon 34113, Korea*
[4] *Max-Planck Institute for Radio Astronomy, Bonn 53121, Germany*
[5] *Korea Research Institute of Standards and Science (KRISS), Daejeon 34113, Korea*
[6] *These authors contributed equally: Minji Hyun and Changmin Ahn*
*e-mail: jungwon.kim@kaist.ac.kr

**Very long baseline interferometry (VLBI) enables high-angular-resolution observations in astronomy and geodesy by synthesizing a virtual telescope with baselines spanning hundreds to thousands of kilometres. Achieving high instrumental phase stability in VLBI relies on the generation of high-quality, atomic-referenced RF local oscillator (LO) and RF-comb signals for the effective downconversion of celestial RF signals and precise phase calibration, respectively. As observing frequencies move into higher ranges with wider bandwidth, conventional electronic methods face significant challenges in maintaining the quality of these signals. Here, we demonstrate that an optical frequency comb (OFC) can be used as a versatile tool to generate and distribute low-noise and atomic-referenced RF-comb and RF-LO signals in the VLBI telescope. Hydrogen maser-stabilized optical pulses are transmitted over a timing-stabilized fibre link from the observatory building to the VLBI receiver system at the telescope, where photodetection converts them into the required RF-comb and RF-LO signals. In VLBI test observation, we successfully detected VLBI fringes and extracted the RF-combs characteristics in a format suitable for VLBI instrumental phase calibration. These**

**results highlight the high potential of OFC-based technology for enhancing next-generation broadband VLBI measurements, advancing astrophysical research and facilitating intercontinental clock comparison.**

**Introduction**

Very long baseline interferometry (VLBI) is a unique measurement technique designed for high angular resolution observations in astronomy and geodesy, which relies on the simultaneous observation of celestial radio sources using distant antennas synchronized to a local atomic clock [1]. By offering an Earth-size virtual telescope that can provide a few tens of micro-arcsecond angular resolution, the VLBI has the potential to reconstruct the shadows of supermassive black holes [2]. The VLBI measures the time delay and spatial coherence of radio frequency (RF) signals of radio sources reaching multiple antennas. These contain not only geometric delay but also other contributions such as atmospheric, on-site instrumental effects, and reference clocks.

While higher frequency VLBI observations enable higher angular resolutions, calibrating the fast fluctuations of water vapor in the troposphere in order to maintain phase coherence of the radio signal coming from the radio source to the antennas has become more difficult. The best solution to correct for this tropospheric effect is simultaneous observations at multiple frequencies. This allows for effective removal of tropospheric delay errors at higher observing frequencies using the measured delays at the lower frequency [3,4]. Furthermore, multi-frequency VLBI observations are a crucial method for studying the physics of black holes and relativistic jets in active galactic nuclei [5,6], stellar evolution [7,8] and geodesy at mm-wavelengths [9].

To fully exploit the superior performance of astronomic and geodetic VLBI at high frequencies, the instrumental phase drifts at each receiver and signal chains must be precisely calibrated. Thus, a phase calibration (PCAL) signal, i.e., a short pulse train corresponding to low-intensity equidistant frequency tones, is injected to provide spectral lines to each receiver band. The extraction of the phase of these tones in the data processing allows for precise measurement of instrumental phase drift and variation throughout the signal chain of a radio telescope during an observation. However, electronics-based PCAL signal generation has a limited frequency range up to 50 GHz without up-converters [10], and it requires complicated equalization due to a high amplitude slope. This can increase the phase uncertainty of PCAL at higher frequency bands due to a weak PCAL signal or destroy the radio signal from a radio source in the sky due to the high amplitude of the PCAL signal at lower frequency bands. The complexities escalate with wideband receivers, of which broad coverage necessitates intricate calibration and equalization techniques.

On the other hand, at VLBI radio telescope receivers, low-noise RF local oscillators (LOs) are used to downconvert the signals coming from the sky. Achieving precise measurements demands the use of highly stable RF LOs that follow the stability of reference clock (frequency standard) at each observatory. As the observation frequency becomes higher for higher resolutions, the generation of higher frequency LO signals with lower phase noise has also become more important. However, the generation of high-frequency microwave signals from the current standard, a hydrogen maser (H-maser), operating at 5 MHz or 10 MHz, mandates the implementation of an intricate multiplication chain, often leading to a compromise in spectral purity. Alternative approaches have been demonstrated using on-site gigahertz-level electronic oscillators and phase-locked loops (PLLs) [11,12]. However, these approaches are only possible to generate a single frequency component at a time, resulting in complexity and bulkiness when applied to multi-frequency receivers demanding multiple LO signals.

Here we show that the employment of an optical frequency comb (OFC) technology inside the VLBI telescopes allows for a simple and high-performance solution to meet both objectives, i.e., distribution and generation of low-noise and atomic reference-locked PCAL and RF LO signals in the VLBI telescope. Mode-locked lasers (MLLs) and OFC sources can generate ultralow-noise optical pulse trains with quantum-limited timing jitter and repetition-rate phase noise [13,14]. With high-speed photodetection, one can generate low-noise photocurrent pulse trains [15] that correspond to the RF comb signals up to the photodetector bandwidth and can be directly utilized for the PCAL signals in VLBI telescopes. At the same time, with subsequent bandpass filtering of this photocurrent pulse train, multiple ultralow-noise single-tone RF signals can be also extracted concurrently, which can then be used for VLBI LO signals. In this work, optical pulse trains locked to the H-maser are delivered from the observatory building to the VLBI antenna via a ~100 m-long timing-stabilized fibre link. At the VLBI antenna, a broadband RF-comb signal with 40 MHz spacing is generated by direct photodetection of delivered pulses up to 50 GHz, and is further injected to the operating PCAL system of the multi-frequency receiver. In addition, 16.64-GHz and 19.2-GHz LO signals were generated, with a measured phase noise level of -126 dBc/Hz at a 100-kHz offset. While the true phase noise is expected to be even lower, it is limited by the noise floor of the measurement instrument. Overall, the phase noise and long-term stability of the signals follow the characteristics of the H-maser source. Note that, while there were latest works on optical link-based time and frequency distribution for radio astronomy including optical carrier, RF frequency, and frequency comb transfer [16-20], this is the first demonstration to transfer the comb signals directly to the antenna receiver room to generate broadband PCAL and ultralow-noise RF LO signals, to the best of our knowledge. Our demonstrated system obtained a successful fringe detection and extraction of PCAL signals at 22 GHz as a first experiment. Furthermore, as the optical frequency comb operates as a coherent gear between optical and microwave frequencies, our approach holds the potential to enhance precision timing in VLBI applications by serving as the last-mile delivery of next-generation optical clocks to the VLBI system, which is critical for intercontinental clock comparison [21,22].

**Results**

**Concept and experimental setup.** Figure 1**a** illustrates the conceptual diagram of the mode-locked fibre laser comb-based system for generating the PCAL and LO signals at the Korean VLBI Network (KVN) Yonsei radio telescope in Seoul, South Korea. A 40-MHz repetition-rate mode-locked fibre laser comb and an H-maser (Kvarz CH1-75A) are placed at the observatory building (H-maser room), while the RF signals generation is conducted in the receiver room at the antenna, denoted by the dashed box in the telescope image (on the right side of Fig. 1**a**). The optical pulse train is transmitted through a ~100-m-long single-mode fibre link to the antenna. At the end of the link, an optical fibre coupler splits the optical pulses into two branches: one is directed toward a Faraday mirror that reflects the signal toward the H-maser room for the link stabilization, while the other branch is used for the RF signals generation for the radio telescope.

The mode-locked fibre laser is synchronized with the local atomic clock, the H-maser, as depicted in Fig. 1**b**. The repetition rate of the laser is set to 40 MHz based on the requirements of the KVN PCAL system. An 800-MHz microwave signal is generated through frequency multiplication of an oven-controlled crystal oscillator (OCXO) phase-locked to the H-maser. The phase difference between this microwave signal and the optical pulse is detected by an electro-optic sampling-based timing detector (EOS-TD) [23-25]. The EOS-TD is a phase-biased Sagnac-loop fibre interferometer that can convert the timing error between the optical pulses and the electric signals (such as sinusoidal RF signals or periodic electric pulse train) into the intensity change of the interferometer outputs. After detecting the EOS-TD outputs using a balanced photodetector (BPD), an error signal proportional to the timing error between the optical pulse and the electric signal can be obtained with sub-fs resolution and drift [25]. The voltage output from the EOS-TD is fed into a proportional-integral (PI) controller, amplified by a high-gain voltage amplifier, and used to control the piezo actuator that adjusts the cavity length of the mode-locked fibre laser. These optical pulses, which deliver the stability of the H-maser, are transferred through a fibre link to the radio telescope antenna for PCAL signal generation (Fig. 1**c**) and LO extraction (Fig. 1**d**) for the multi-band VLBI receiver.

To compensate for various environmental fluctuations imposed on the ~100-m-long fibre link between the H-maser room and the antenna room, such as mechanical stress, acoustic noise and temperature loads, another EOS-TD is employed to measure and compensate for the fibre delay error (Fig. 1**e**). Compared to the well-established optical cross-correlation technique [26-29], the main advantages of using the EOS-TD for link stabilization is that it can detect the timing error between the optical pulses and the electric signals with sub-fs resolution even when the optical pulsewidth is much longer than ps [30], therefore does not require strict dispersion compensation in the optical pulse distribution. The use of electric signals for measuring the optical pulse position also allows for much wider timing detection range of more than tens to hundreds ps.

After amplification, optical pulses from the mode-locked fibre laser comb are split into two streams: one as a reference signal for the EOS-TD and the other (red pulses in Fig. 1**e**) sent through a dispersion-compensating fibre and fibre stretcher for length compensation before transmission to the antenna. The reflected pulses (blue pulses in Fig. 1**e**) are converted into an electronic signal, and applied to the EOS-TD. Two RF signals—either harmonics from photocurrent pulses or a 1.4 GHz microwave signal—can be used for timing stabilization, with feedback provided by a PI controller and a high-gain voltage amplifier to the fibre stretcher for accurate fibre link timing.

Through the timing-stabilized fibre link, optical pulses are converted to low-jitter photocurrent pulses via a high-linearity photodiode at the antenna, serving as a PCAL signal for the VLBI system with low optical power (300 µW) to prevent interference with astronomical signals, while still high enough to extract the PCAL signal with a sufficient signal-to-noise ratio. The same photocurrent pulses are used to generate a low-phase-noise microwave signal for the LO, where pulsed illumination provides a lower noise floor and higher power efficiency. To boost the microwave signal, a five-stage fibre-based pulse repetition-rate multiplier [31,32] increases the repetition rate to 1.28 GHz, enhancing power at the 13th harmonic (16.64 GHz) and 15th harmonic (19.2 GHz) for the LO signal, with harmonics filtered and amplified to meet receiver power requirements. More detailed information on the experiment setup can be found in Methods.

**H-maser-comb synchronization.** Figure 2 shows the phase noise spectra related to H-maser-comb synchronization. Curve (i) shows the repetition-rate phase noise of the free-running 40-MHz fibre comb laser, when scaled to 800-MHz carrier frequency (20th harmonic). Curve (ii) represents the absolute phase noise of an 800 MHz microwave signal generated from the OCXO phase-locked to the H-maser, measured by the signal source analyser (SSA). For comparison, the absolute phase noise of 5-MHz H-maser output (scaled to 800-MHz carrier frequency) is also shown by curve (iii), obtained from the manufacturer's datasheet. The residual phase noise for H-maser-comb synchronization, measured by the out-of-loop EOS-TD, is presented by curve (iv) (see Methods). These results show that the residual phase noise in the H-maser-comb synchronization remains sufficiently low compared to the H-maser noise, indicating effective phase stability transfer of the H-maser to the antenna room via optical pulse trains.

**Fibre link stabilization.** To guarantee complete transfer of H-maser stability, the residual noise of the fibre link should be far lower than that of H-maser. The measured residual timing jitter spectral density of the fibre link stabilization is presented in Fig. 3**a**. For comparison, the phase noise of the H-maser, provided in the manufacturer's test sheet, is also included. Curve (i) shows the residual jitter density when the photocurrent pulses are used for the RF input to the EOS-TD. The residual jitter density is $4.6 \times 10^{-4}$ fs$^2$/Hz, $2.4 \times 10^{-5}$ fs$^2$/Hz, and $7 \times 10^{-6}$ fs$^2$/Hz at 10-Hz, 1-kHz, and 1-MHz frequency, respectively. Curve (ii) shows the phase noise when an extracted 1.4-GHz microwave signal is used for the RF input to the EOS-TD. The residual jitter density at 10-Hz, 1-kHz, and 1-MHz offset is $3 \times 10^{-3}$ fs$^2$/Hz, $8 \times 10^{-4}$ fs$^2$/Hz, and $6 \times 10^{-4}$ fs$^2$/Hz, respectively. The integrated rms timing jitters using photocurrent pulses and microwave signal are 2.6 fs and 24.5 fs integrated from 1 Hz to 1 MHz, respectively. Figure 3**b** shows the long-term residual timing drift measurement for both methods, with the rms timing drifts over 1,000 s of 1.4 fs (photocurrent pulse) and 2.6 fs (microwave signal). This fibre link stabilization system can operate over 30,000 s without loss of coherence, exceeding typical VLBI observation scan durations under 1,000 s. The computed fractional frequency instability based on this timing drift measurement is shown in Fig. 3**c**, starting from $1.2 \times 10^{-14}$ ($9.6 \times 10^{-14}$) at 0.1 s averaging time and reaches to $<6.4 \times 10^{-18}$ ($<1.2 \times 10^{-17}$) at 1,000 s averaging time for photocurrent pulses (microwave signal). The photocurrent pulse-based stabilization achieves better performance due to higher detection sensitivity and reduced relative jitter, and it simplifies the setup by eliminating RF amplifiers and bandpass filters, thereby reducing environmental noise. Given that both methods produce residual noise levels significantly below H-maser phase noise, either configuration with EOS-TD is suitable for link stabilization.

**RF-comb signal generation.** The RF spectrum of the RF-comb signals generated in the antenna room was measured using a 50-GHz bandwidth RF spectrum analyser (Keysight,

N9030A) with a resolution bandwidth of 51 kHz. It contains integer harmonics of repetition rate within the photodiode bandwidth, shown as normalized amplitude in Fig. 4**a**. The used RF cable (Gore, Phaseflex) has frequency-dependent insertion loss as plotted by a dashed line in Fig. 4**a**, indicating that the RF-comb amplitude remains nearly uniform up to 50 GHz once the cable loss is accounted for. Note that the frequency range of RF-comb can be extended beyond 100 GHz with the higher bandwidth photodiodes [33], which would not be accomplished by conventional electrical RF-comb generators. The power spectrum of RF-comb signal at K-band injected to the VLBI receiver is shown in Fig. 4**b**. Each frequency component is equal-distanced by repetition rate of optical pulse, 40 MHz in this case, covering the entire K-band as required for the PCAL signal for the KVN-Yonsei telescope.

**RF-LO signal generation.** Figure 5 shows the absolute phase noise performance of the extracted RF-LO signals measured in the antenna receiver room by a signal-source analyser (SSA). The broad peak around 3 kHz is the servo bump resulting from the H maser-comb synchronization PLL (Fig. 2). Free-running laser noise is measured using the same SSA by O-E conversion and extracting the fundamental harmonic of laser, then rescaled to carrier frequency of corresponding LO. Within the locking bandwidth, the phase noise of the 16.64-GHz (19.2-GHz) signal closely follows the phase noise of the H-maser, indicating a complete transfer of frequency stability. It's worth noting that the significant spurious noises ranging from 100 Hz to 1 kHz originate from the H-maser, affecting not only the H-maser-laser stability transfer loop but also the phase noise of the LO itself. The SSB phase noise reaches -126 dBc/Hz (-124 dBc/Hz) at a 100 kHz offset frequency. Note that these results are limited by the instrumental noise of SSA in the offset frequency range of <10 Hz and >200 kHz. Given the input optical power to the photodiode, the thermal noise-limited noise floor [34] is projected to be -149 dBc/Hz (-148 dBc/Hz) at 16.64 GHz (19.2 GHz), and with the RF amplifier noise figure included, it will reach -146 dBc/Hz (-145 dBc/Hz). These results indicate that the actual RF-LO phase noise is considerably lower than the measured result, which is limited by the SSA instrument noise. For the operation of KVN-Yonsei, photonic RF-LO signal at 16.64 GHz was injected to the VLBI receiver.

**VLBI observation and PCAL extraction.** The first VLBI experiment using the KVN-Yonsei radio telescope, integrated with photonic-generated PCAL and LO signal chains, was conducted from 21:48 UTC on 14 May 2024 to 02:05 UTC on 15 May 2024. All four KVN radio telescopes – Yonsei (KY), Ulsan (KU), Tamna (KT), and Pyeongchang (KC) – participated, observing 40 well-known VLBI sources in 2-minute scans per source. A simultaneous dual-frequency setup at 22 GHz and 43 GHz was employed with multi-frequency receivers in both right-hand circular polarization (RHCP) and left-hand circular polarization (LHCP), each with a 512-MHz bandwidth (512 MHz × 4 channels), resulting in an 8-Gbps data rate. Photonic PCAL signals were injected into the RHC coupling port of the 22-GHz receiver at KY. The interferometric correlation was processed using the DiFX software correlator [34] with a 1-s integration time and 62.5-kHz frequency sampling. Fringe fitting was performed using the NRAO Astronomical Image Processing System (AIPS), with KU as the reference antenna. The fringe detection at KY showed a signal-to-noise ratio (SNR) ranging from 20 to 350, with an overall detection rate of approximately 92% across frequency and polarization bands, indicating stable performance of the photonic RF-LO throughout the observation. The PCAL tone extraction was performed, confirming stable amplitudes and phases with minimal phase excursions during the observation period (as shown in Figs. 6 and 7). The results demonstrate the robustness of photonic RF signal generation and distribution, validating its effectiveness for VLBI phase calibration and high-frequency observation.

**Discussions**

In summary, we have proposed and demonstrated the OFC-based RF signal distribution and generation, offering low-phase noise signals for multi-frequency VLBI receivers. The current demonstration involves transmitting an OFC signal to the antenna, extracting the PCAL and LO signals, and then injecting them into the VLBI receiver. At present, the frequency range of PCAL signal generation has been demonstrated up to 50 GHz, limited by the frequency range of the RF analyser used. The frequency range of the PCAL signal would be determined by the bandwidth of the optical signal and the capabilities of the photodiode. Given the broad bandwidth of optical signals, the bandwidth of the photodiode would set the upper limits, with commercial parts exceeding 100 GHz readily available. Therefore, expanding the frequency range of the proposed system for mm-wavelength VLBI applications is feasible. For example, the KVN-style multi-frequency receiving system and the compact triple-band receiving (CTR) system [36] are now being globally introduced to existing VLBI radio telescopes within the European VLBI Network (EVN) and the Global mm-VLBI Array (GMVA), whose observing frequencies range from 18 – 116 GHz. These provide new opportunities for ultra-precise astrometry and geodesy in mm-VLBI by calibrating two major sources of errors: atmospheric phase fluctuation errors, as widely demonstrated in the previous works [37-39], and instrumental phase errors, as demonstrated in this study.

It is worth noting that since the photocurrent pulse contains numerous frequency components, multiple microwave signals at different frequencies, multiples of the effective repetition rate, can be extracted for other observation frequency bands. For instance, a 34.56 GHz microwave can be extracted and used as an LO signal for the second observation frequency band of the KVN system. Consequently, in conjunction with the simultaneous multi-frequency receiving system, our demonstrated system has the potential to surpass frequency thresholds for precision astrometric measurements.

While the current demonstration relied on an H-maser, the local atomic clock used in the KVN VLBI system, our OFC-based approach has the capability to coherently link optical frequencies to the microwave frequency domain. This approach can seamlessly complete the final step necessary for transferring optical atomic clock stability to the actual receivers in the VLBI system. This is particularly significant given the ongoing efforts to enhance clock references using higher stability frequency standards, such as the optical atomic clock disseminated via a global network [16, 17]. Furthermore, the capability of high-resolution astronomical observations, high-frequency broadband VLBI is expected to be used for intercontinental comparisons of remote optical clocks with unprecedented accuracy because high-frequency broadband observation can reduce the source structure effect, a dominant error in the previous experiment [21]. For this purpose, it is vital to suppress other instrumental errors over large width of the frequency band, and this can be accomplished by photonic-generated PCAL and LO signals.

**Methods**

**Mode-locked fibre oscillator-based optical frequency comb source.** The mode-locked laser is constructed using an all-polarization-maintaining (PM) fibre-based nonlinear amplifying loop mirror (NALM) in a figure-of-nine configuration [40,41]. To enable self-starting, a nonreciprocal phase bias – comprising a half-wave plate, Faraday rotator, and octave-wave plate – is incorporated into the linear arm. The cavity length is adjusted to set the laser's repetition rate at 40 MHz, which is the RF-comb spacing for the PCAL signal generation, and a PZT actuator attached to the mirror in the linear arm allows for modulation of this repetition rate.

**Fibre link stabilization.** Optical pulses from the mode-locked fibre laser comb are amplified by an Erbium-doped fibre amplifier (EDFA) and then split into two streams. One stream serves as a reference signal, which is utilized for the optical input of the EOS-TD. The other stream passes dispersion compensating fibre (DCF) to mitigate fibre link dispersion, then proceeds through a circulator and a fibre stretcher, which acts as an actuator for fibre length compensation, before being transmitted via the fibre link to the antenna. The reflected pulses from the end of the fibre link are photodetected by a 12-GHz p-i-n photodiode. For the RF input to the EOS-TD, two different electronic signals can be used; one is photocurrent pulses and the other is the single-frequency (1.4 GHz = 35 × 40 MHz) microwave signal. The reference optical pulses are adjusted to be located at the middle of the rising edge of photocurrent pulses, which has an attosecond-level relative jitter [15], in the first case and the zero-crossing of the microwave signal in the second case, as shown in the inset of Fig. 1**e**. The error signal from the EOS-TD is then fed back to the fibre stretcher through a proportional-integral (PI) controller and a high-gain voltage amplifier for timing stabilization of the fibre link. To assess the link stabilization performance, an out-of-loop EOS-TD is used. Reference optical pulses are split into two paths by an optical coupler, while electronic signals are divided by a power splitter. These optical and electronic signals are directed to in-loop EOS-TD for link stabilization and to out-of-loop EOS-TD for residual noise measurement [42]. The residual jitter spectral density is calculated by converting the voltage spectrum to timing jitter using the EOS-TD sensitivity. Measurements are conducted with a fast Fourier transform (FFT) analyser (Stanford Research Systems, SR770) for 1 Hz - 100 kHz and an RF spectrum analyser (Agilent, E4411B) for 100 kHz - 1 MHz offset frequency ranges.

**RF-comb signal generation.** Through the timing-stabilized fibre link, optical pulses arriving at the antenna receiver room are converted to photocurrent pulses through a high-linearity photodiode. Low-jitter photocurrent pulses are equivalent to the broadband RF comb in the frequency domain. Thereby, it can serve as a PCAL signal for the VLBI system. A low optical power of 300 µW is applied to prevent disturbance in signals from astronomical radio sources, but to extract the PCAL signal with a sufficient signal-to-noise ratio. This low incident power would be beneficial to avoid photodiode saturation.

**RF-LO generation.** To achieve higher microwave signal power and to alleviate saturation effect for the LO, fibre-based optical pulse repetition rate multiplier [31,32] is implemented. Five-stage cascaded multiplier gives effective repetition rate of 1.28 GHz (=40 MHz × $2^5$). This results in 21-dB power enhancement at 16.64 GHz (13-th harmonic of 1.28 GHz), when compared to the 416th harmonic of the fundamental repetition rate of 40 MHz. Rate-multiplied optical pulses are applied to a high-speed photodiode (Freedom Photonics, FP1015C) at bias voltage of 8 V and photocurrent of 1 mA. Then, 13-th harmonic is selected through bandpass filtering followed by an RF amplifier to match the power requirement of receiver. In addition to 13-th harmonic, the 15-th harmonic (19.2 GHz) was also obtained by changing the bandpass filter centre frequency to show multiple frequency generation capability. For the operation of KVN-Yonsei, 16.64-GHz LO signal was injected to the VLBI receiver.

**PCAL signal extraction.** The correlator combines the basebands of all telescopes to computationally form time averaged interferometric data - measurements of the spatial coherence of the astronomical source as seen by telescope-pairs, related to the Fourier transformed image of the source. Apart of interferometric data, the DiFX correlator can also output calibration data recovered from signals in the baseband data (e.g., switched-noise power, PCAL tones) and export them in a format readable by most VLBI postprocessing software. The PCAL tone extractor in DiFX is designed for standard geodetic and astronomical PCAL systems. With our photonic PCAL system, we were able to achieve a VLBI-friendly tone comb setup having a spacing of 40 MHz, combined with normal VLBI receiver and the digitized baseband recordings that had 2-bit quantization.

## Data availability

The data that support the findings of this study are available from the corresponding authors upon request.

## Code availability

The simulation and computational codes for this study are available from the corresponding authors on request.

**Acknowledgements**

This research was supported by the National Research Council of Science and Technology (NST) of Korea (Grant CAP22061-000), the National Research Foundation (NRF) of Korea (Grants RS-2024-00334727, RS-2024-00436737, RS-2021-NR060086), and the Institute for Information and Communications Technology Promotion (IITP) of Korea (Grant RS-2023-00223497). The KVN and a high-performance computing cluster are facilities operated by the Korea Astronomy and Space Science Institute (KASI). The KVN observations and correlations are supported through the high-speed network connections among the KVN sites provided by the Korea Research Environment Open NETwork (KREONET), which is managed and operated by the Korea Institute of Science and Technology Information (KISTI).

**Author contributions**

J.K., T.J. and M.-S.H. conceived the idea and supervised the project. M.H., C.A. and J.K. designed the experimental setup. M.H., C.A., J.C., J.B., W.J., D.-H.J. and D.-Y.B. built and installed the system at KVN-Yonsei and obtained the data. M.H., C.A., T.J., J.W. and J.K. analysed the data. M.H., T.J. and J.K. wrote the manuscript with inputs from all authors.

**Competing interests**

J.K., M.-S.H., T.J., D.-H.J., M.H. and C.A. are inventors on a patent application related to this work filed by KAIST, KRISS and KASI (Korean patent application 10-2024-0050794 filed on April 16, 2024). The authors declare no other competing interests.

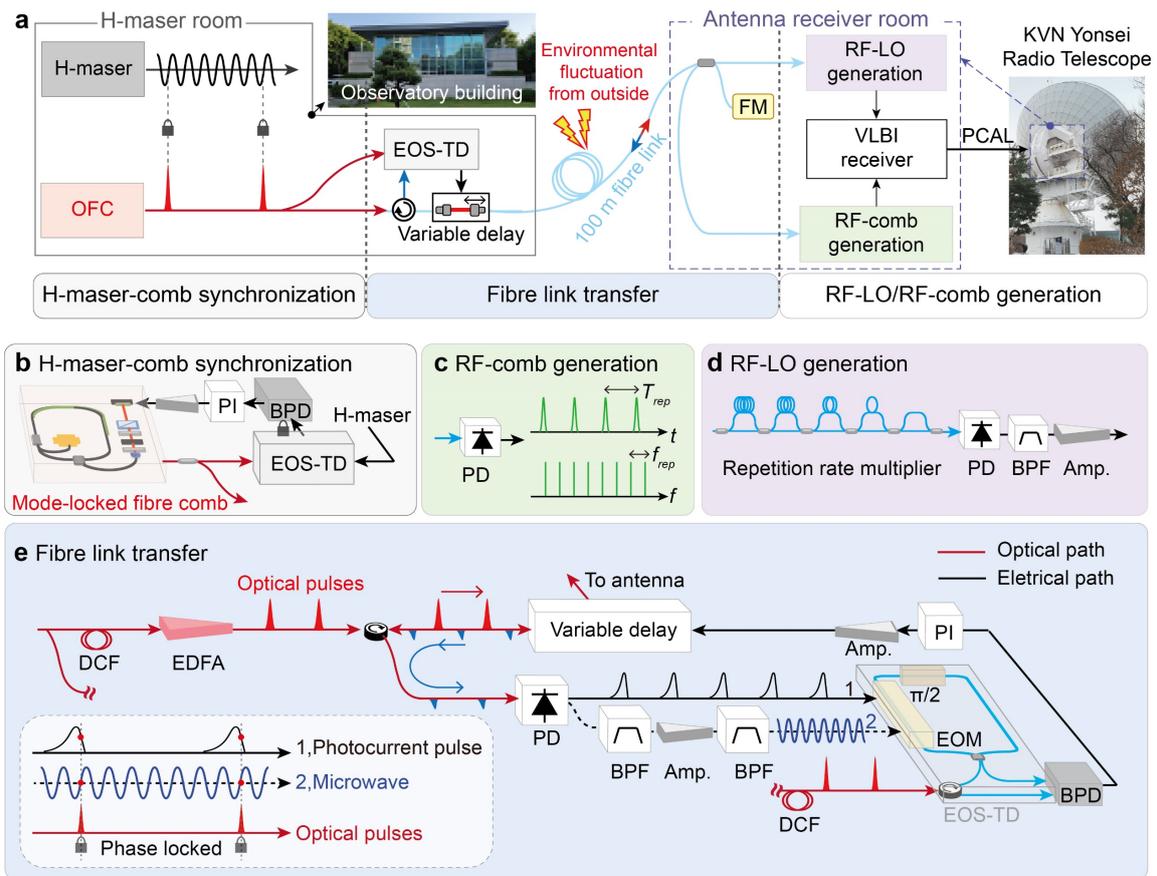

**Fig. 1 Photonic generation and distribution of RF signals at VLBI radio telescope. a,** Schematic of the demonstrated photonic system that delivers H-maser-locked optical pulses from the observatory building to the antenna room to generate RF-LO and PCAL signals for VLBI. **b,** H maser-comb synchronization setup. BPD, balanced photodetector; PI, proportional-integral servo controller. **c,** RF-comb and PCAL signal generation setup. **d,** RF-LO extraction setup. **e,** Link stabilization setup using the EOS-TD. Two different methods ((1) photocurrent pulse and (2) microwave) have been demonstrated. DCF, dispersion compensating fibre; EDFA, Erbium-doped fibre amplifier; EOM, Electro-optic phase modulator, OFC, optical frequency comb source.

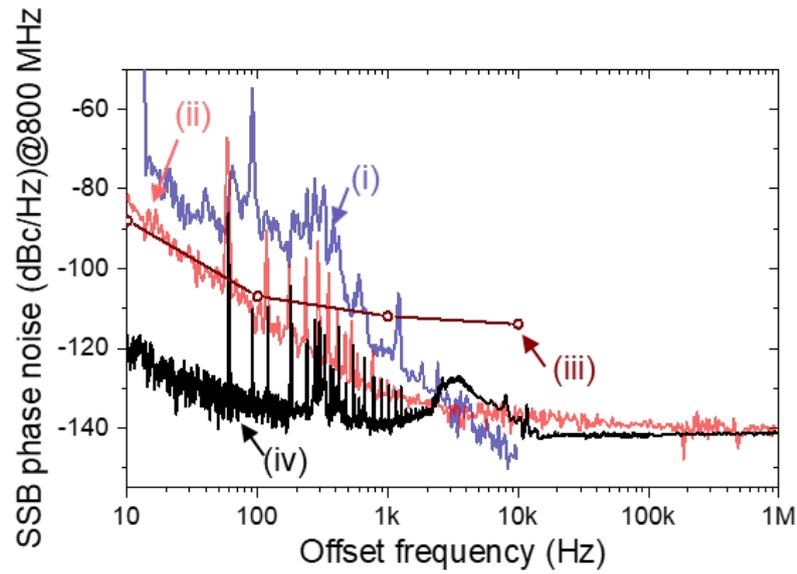

**Fig 2. H-maser-comb synchronization performance.** (i) Absolute repetition-rate phase noise of the free-running 40-MHz fibre comb laser (scaled to 800-MHz carrier frequency). (ii) Absolute phase noise of the 800 MHz microwave signal phase-locked to the H-maser when measured by the SSA. (iii) Absolute phase noise of 5-MHz H-maser signal obtained from manufacturer's test sheet. (iv) Residual phase noise for H-maser-comb synchronization when measured by the out-of-loop EOS-TD.

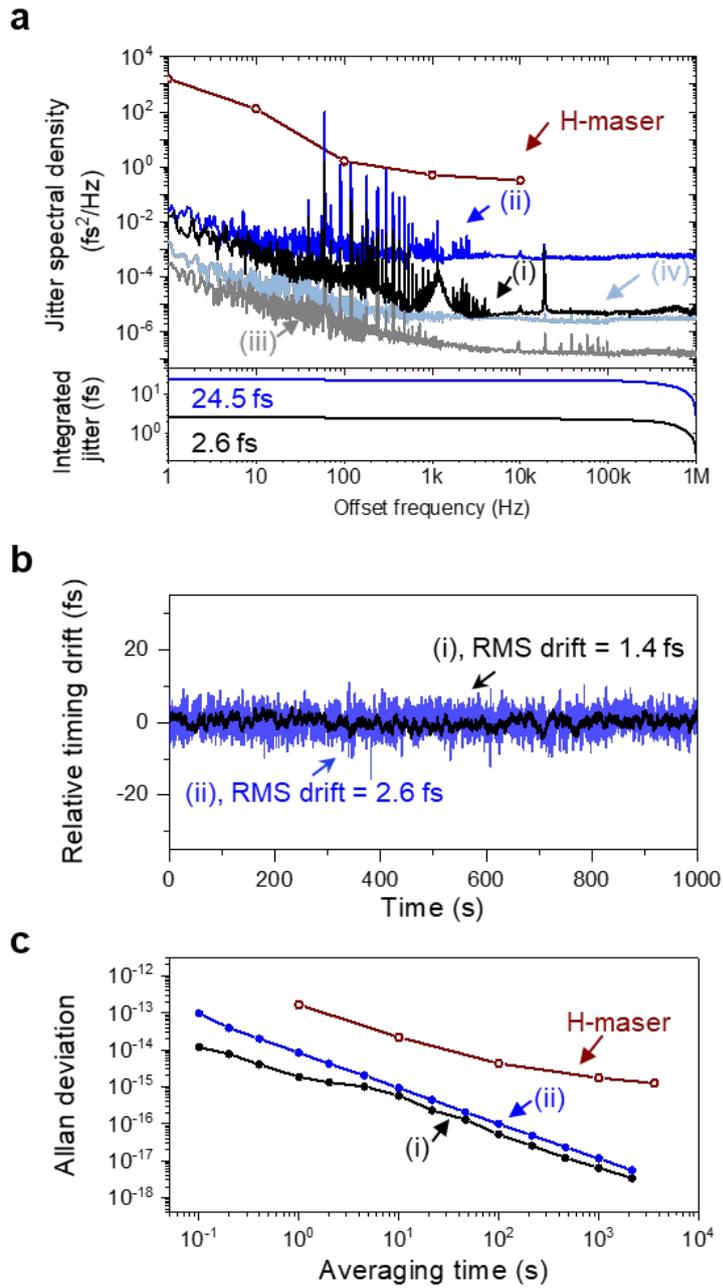

**Figure 3. Link stabilization performance** **a** Residual jitter spectral density using (i) photocurrent pulses and (ii) extracted microwave, their measurement backgrounds (iii), (iv), respectively. **b** Long-term residual timing drift. **c** Allan deviation versus averaging time using (i) photocurrent pulses and (ii) extracted microwave. For comparison, H-maser performance from manufacturer's test sheet is presented.

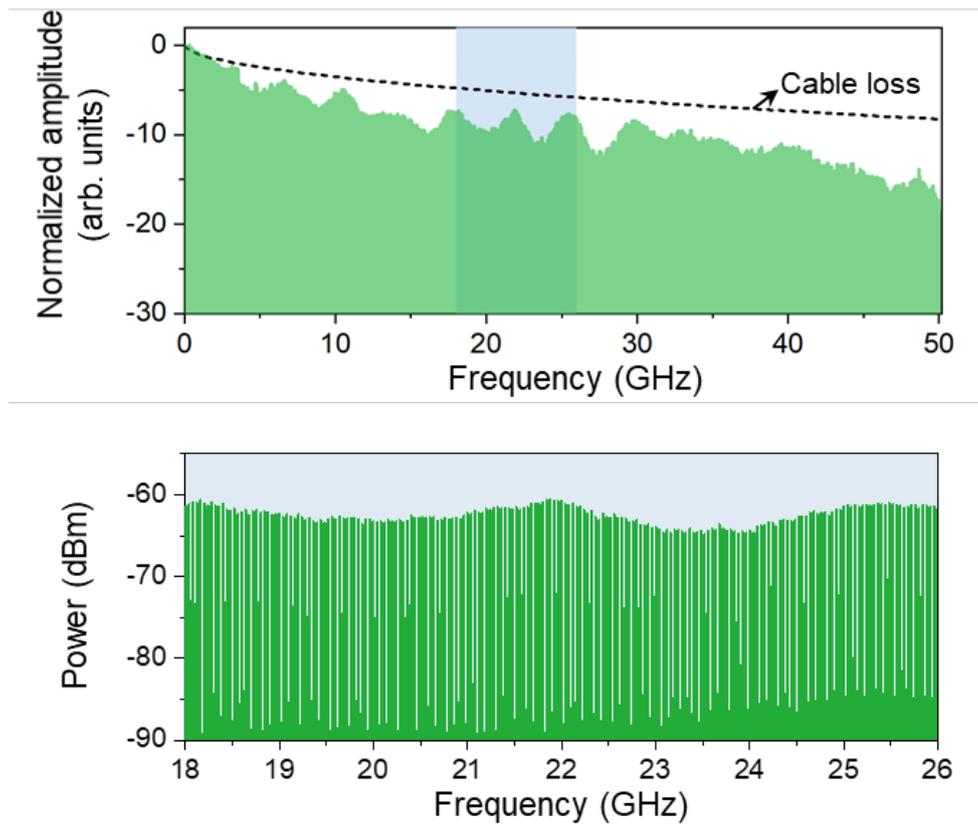

**Figure 4. RF-comb signal generation results. a,** Generated RF-comb signal up to 50 GHz presented in normalized amplitude. **b,** Power spectrum of RF-comb signal at K-band. The resolution bandwidth is set to 51 kHz.

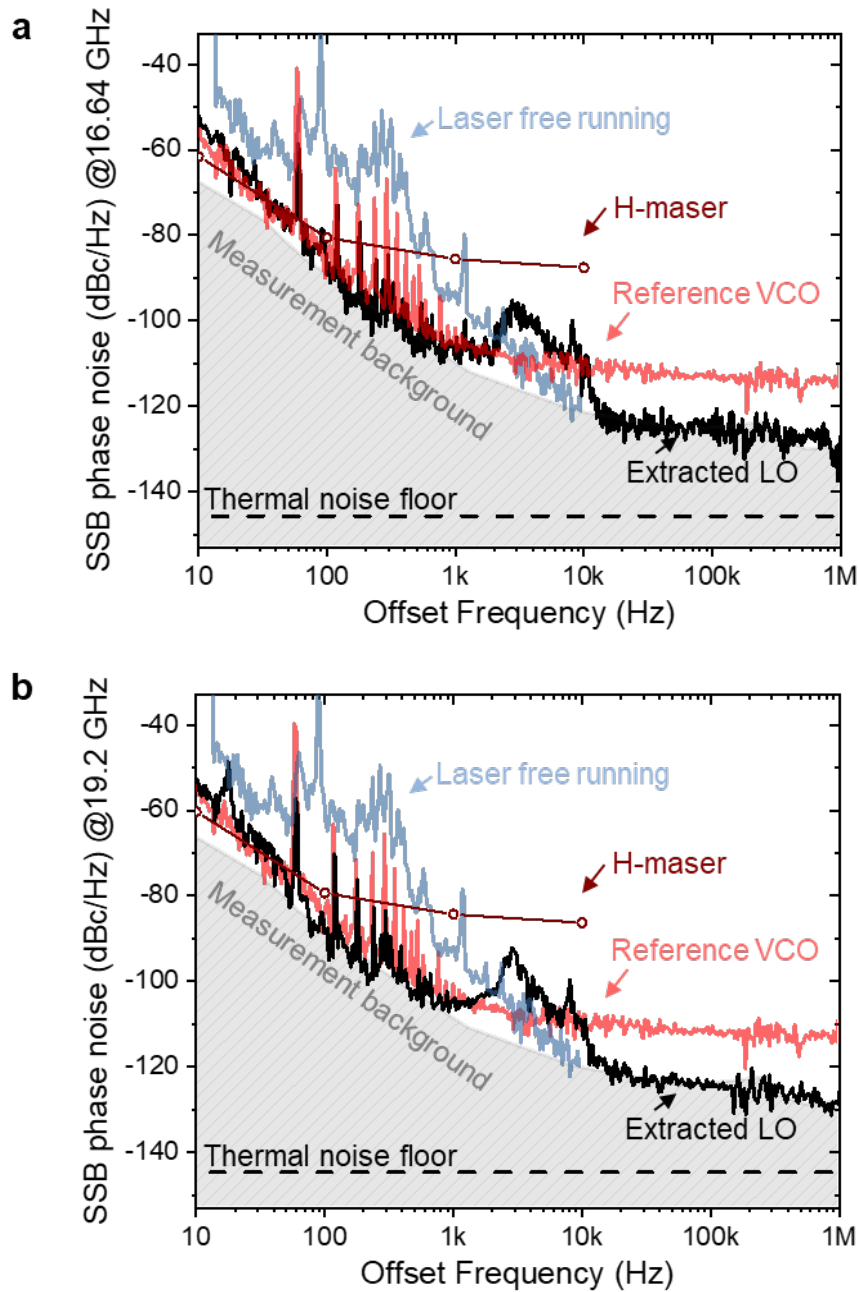

**Figure 5. RF-LO signal phase noise performance.** Measured single-sideband (SSB) absolute phase noise performances of the extracted LOs at (**a**) 16.64 GHz and (**b**) 19.2 GHz. The thermal noise floor is a computed from the input optical power to the photodiode and the RF amplifier noise figure.

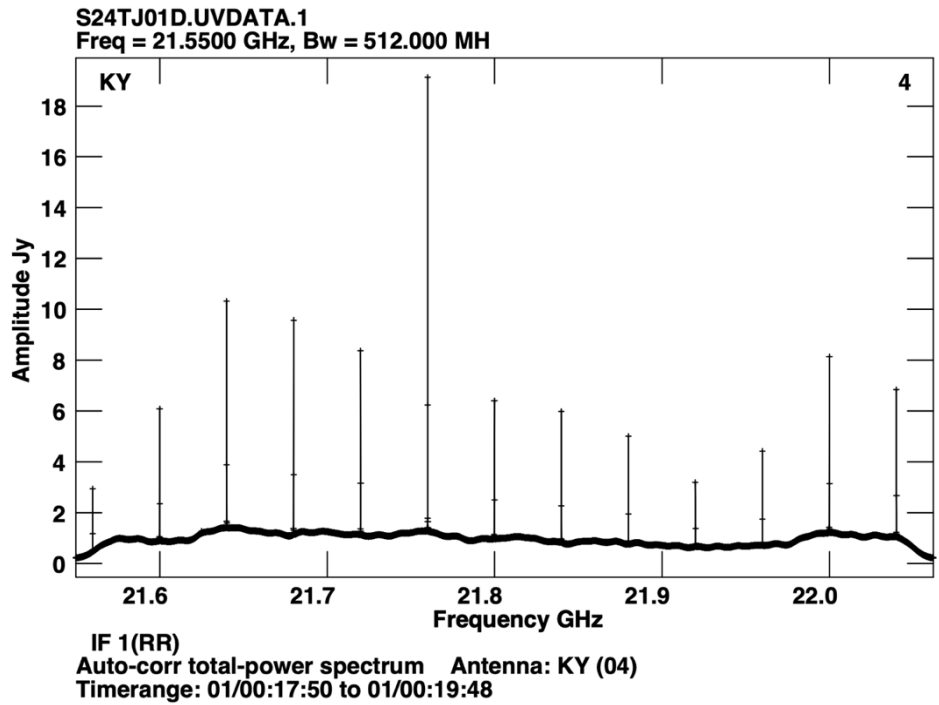

**Figure 6. Power spectrum of PCAL tones.** Power spectrum of PCAL tones observed at the KVN Yonsei (KY) radio telescope, showing 13 distinct tones within a 512 MHz bandwidth centered at 21.086 GHz (RHCP). Each tone is spaced at intervals of 40 MHz.

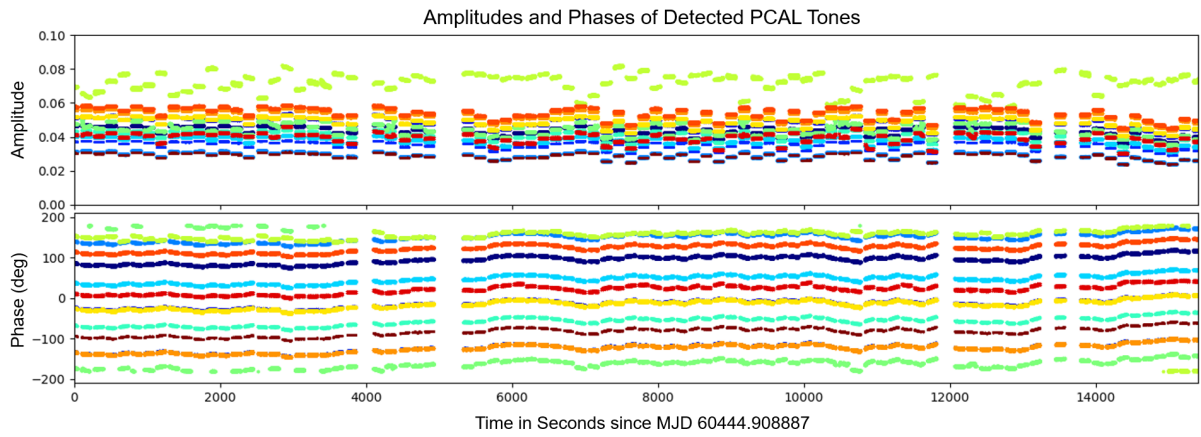

**Figure 7. Amplitude and phase of detected PCAL tones at KY during the observation.**
Total 13 tones with 40-MHz spacing were presented in the bandwidth of 512 MHz at 22 GHz (RHCP). The frequencies of tones are 22040, 22000, 21960, 21920, 21880, 21840, 21800, 21760, 21720, 21680, 21640, 21600, 21560 MHz in reversed order due to lower side band.